\documentclass[longauth]{aa} 
\usepackage{graphicx}
\usepackage{txfonts}
\begin{document}
   \title{The Herschel-ATLAS: Extragalactic Number Counts from 250 to 500 Microns\thanks{Herschel is an ESA space observatory with science instruments provided by European-led Principal Investigator consortia and with important participation from NASA.}}

\author{D.L. Clements\inst{1}
\and
E. Rigby\inst{2}
\and
S. Maddox\inst{2}
\and
L. Dunne\inst{2}
\and
A. Mortier\inst{1}
\and
C. Pearson\inst{19, 20}
\and
A. Amblard\inst{4}
\and
R. Auld\inst{5}
\and
M. Baes\inst{6}
\and 
D. Bonfield\inst{4}
\and
D. Burgarella\inst{8}
\and
S. Buttiglione\inst{9}
\and
A. Cava\inst{10}
\and
A. Cooray\inst{4}
\and
A. Dariush\inst{5}
\and
G. de Zotti\inst{9}
\and
S. Dye\inst{5}
\and
S. Eales\inst{5}
\and
D. Frayer\inst{11}
\and
J. Fritz\inst{6}
\and
Jonathan P. Gardner\inst{22}
\and
J. Gonzalez-Nuevo\inst{13}
\and
D. Herranz\inst{13}
\and
E. Ibar\inst{3}
\and
R. Ivison\inst{3}
\and
M.J. Jarvis\inst{7}
\and
G. Lagache\inst{14}
\and
L. Leeuw\inst{15}
\and
M. Lopez-Caniego\inst{13}
\and
M. Negrello\inst{16}
\and
E. Pascale\inst{5}
\and
M. Pohlen\inst{5}
\and
G. Rodighiero\inst{9}
\and
S. Samui\inst{12}
\and
S. Serjeant\inst{16}
\and
B. Sibthorpe\inst{3}
\and
Douglas Scott\inst{22}
\and
D.J.B. Smith\inst{5}
\and
P. Temi\inst{15}
\and
M. Thompson\inst{7}
\and
I. Valtchanov\inst{17}
\and
P. van der Werf\inst{21}
\and
A. Verma\inst{18}
}

\institute{
Astrophysics Group, Imperial College, Blackett Laboratory, Prince
Consort Road, London SW7 2AZ, UK
\and
School of Physics and Astronomy, University of Nottingham,
University Park, Nottingham NG7 2RD, UK
\and
UK Astronomy Technology Center, Royal Observatory Edinburgh, Edinburgh, EH9 3HJ, UK
\and
Dept. of Physics \& Astronomy, University of California, Irvine, CA 92697, USA
\and
School of Physics and Astronomy, Cardiff University,
  The Parade, Cardiff, CF24 3AA, UK
\and
Sterrenkundig Observatorium, Universiteit Gent, Krijgslaan 281 S9,
B-9000 Gent, Belgium
\and
Centre for Astrophysics Research, Science and Technology Research
Centre, University of Hertfordshire, Herts AL10 9AB, UK
\and
Laboratoire d'Astrophysique de Marseille, UMR6110 CNRS, 38 rue F.
Joliot-Curie, F-13388 Marseille France
\and
University of Padova, Department of Astronomy, Vicolo Osservatorio
3, I-35122 Padova, Italy
\and
Instituto de Astrof\'{i}sica de Canarias, C/V\'{i}a L\'{a}ctea s/n, E-38200 La
Laguna, Spain
\and
National Radio Astronomy Observatory,  PO Box 2, Green Bank, WV  24944, USA
\and
Scuola Internazionale Superiore di Studi Avanzati, via Beirut 2-4,
34151 Triest, Italy
\and
Instituto de F\'isica de Cantabria (CSIC-UC), Santander, 39005, Spain
\and
Institut d'Astrophysique Spatiale (IAS), Bâtiment 121, F-91405 Orsay, France; and Université Paris-Sud 11 and CNRS (UMR 8617), France
\and
Astrophysics Branch, NASA Ames Research Center, Mail Stop 245-6, Moffett Field, CA 94035, USA
\and
Dept. of Physics and Astronomy, The Open University, Milton Keynes, MK7 6AA, UK
\and
Herschel Science Centre, ESAC, ESA, PO Box 78, Villanueva de la
Ca\~nada, 28691 Madrid, Spain
\and
Oxford Astrophysics, Denys Wilkinson Building, University of Oxford, Keble Road, Oxford, OX1 3RH
\and
SSTD, Rutherford Appleton Laboratory, Chilton, Didcot, Oxfordshire OX11 0QX, UK
\and
Institute for Space Imaging Science
University of Lethbridge,
Lethbridge, Alberta CANADA, T1K 3M4
\and
Leiden Observatory, Leiden University, PO Box 9513, NL - 2300 RA Leiden, The Netherlands
\and
Astrophysics Science Division, Observational Cosmology Laboratory, Code 665, Goddard Space Flight Center, Greenbelt MD 20771, USA
\and
Department of Physics \& Astronomy, University of British Columbia, 6224 Agricultural Road, Vancouver, B.C., V6T 1Z1, Canada
}

   \date{}

 

\abstract
{}
 {The Herschel-ATLAS survey (H-ATLAS) will be the largest area survey to be undertaken by the Herschel Space Observatory. It will cover 550 sq. deg. of extragalactic sky at wavelengths of 100, 160, 250, 350 and 500$\mu$m when completed, reaching flux limits (5$\sigma$) from 32 to 145mJy. We here present galaxy number counts obtained for SPIRE observations of the first $\sim$14 sq. deg. observed at 250, 350 and 500$\mu$m.}
{Number counts are a fundamental tool in constraining models of galaxy evolution. We use source catalogs extracted from the H-ATLAS maps as the basis for such an analysis. Correction factors for completeness and flux boosting are derived by applying our extraction method to model catalogs and then applied to the raw observational counts. }
 {We find a steep rise in the number counts at flux levels of 100---200mJy in all three SPIRE bands, consistent with results from BLAST. The counts are compared to a range of galaxy evolution models. None of the current models is an ideal fit to the data but all ascribe the steep rise to a population of luminous, rapidly evolving dusty galaxies at moderate to high redshift.
 }
{}

   \keywords{Galaxies: evolution; Galaxies: statistics; Infrared: galaxies; Submillimetre: galaxies
               }

\authorrunning{D.L. Clements et al.}
   \maketitle
%

\section{Introduction}

The determination of the number of sources observed as a function of brightness is fundamental in observational astronomy. These so-called number counts have been used in the past to determine the structure of our Galaxy (Herschel, 1784), rule out cosmological models (Ryle \& Clark, 1961), and examine the evolution of galaxies and quasars (eg. Smail et al., 1995; Hasinger et al., 1993). Number counts have revealed that very rapid evolution must take place among infrared luminous populations to account for the large number of sources seen at faint fluxes in the submillimetre (eg. Hughes et al., 1998; Smail et al., 1997; Eales et al., 2000) and to produce the cosmic infrared background (Puget et al., 1996; Fixsen et al., 1998; Hauser \& Dwek 2001). The detailed origin and nature of this evolution remains unclear following observations in the mid-IR (eg. Elbaz et al., 1999, Papovich et al., 2004, Le Floc'h et al., 2009; Babbedge et al., 2006) and far-IR (eg. Frayer et al., 2006) using {\em ISO} and {\em Spitzer}, and extensive followup studies of submm sources (eg. Chapman et al., 2005; Clements et al., 2008; Dye et al., 2008). Issues needing to be addressed include spectral energy distribution (SED) evolution and the redshift distribution of the evolving luminous populations. Understanding these processes is a key goal of the Herschel Space Observatory (Pilbratt et al., 2010) and of the various large extragalactic surveys that it is undertaking. We here report the first number counts obtained by the Herschel-ATLAS (H-ATLAS) survey (Eales et al., 2010) which set new constraints on models of galaxy evolution in the far-IR/submm.


\section{Observations and Source Extraction}

The overall H-ATLAS survey is described by Eales et al. (2010). Briefly, it is a survey that will cover $\sim$550 sq. deg. of extragalactic sky at wavelengths of 100, 160, 250, 350 and 500$\mu$m to 5$\sigma$ depths of 32-145mJy. The observations discussed here cover a $\sim$14 sq. deg. region, $<$3\% of the total area, which were obtained during the Science Demonstration Phase (SDP) of Herschel operations. This field is centred on RA=09:05:30.0 DEC=00:30:00.0. The field lies in the centre of one of the Galaxy And Mass Assembly (GAMA) optical survey fields (Driver et al., 2009), GAMA9, so plentiful complementary data from UV to near-IR, including optical spectra, is available.

H-ATLAS uses parallel mode observations which provide data simultaneously from both the PACS (Poglitsch et
al. 2010) and SPIRE (Griffin et al. 2010) instruments.
The time-line data is reduced using a naive mapping technique after removing
instrumental temperature variations from the time-line data (Pascale
et al. in prep., see also Griffin et al., 2008). Noise maps were generated using the two
cross-scan measurements to estimate the noise per detector pass, and
then for each pixel the noise is scaled by sqrt(number of detector
passes). PACS data and results will be discussed elsewhere.

Sources were identified in the SPIRE maps by first subtracting a local
background, estimated from the peak of the histogram of pixel values
in $30\times 30$ blocks, corresponding to 2.5' for the 250$\mu$m
map, and 5' for the 350 and 500$\mu$m maps. Larger blocks, with $60\times 60$ and $120\times 120$ pixels were tried, but these produced no
significant differences ($> \sim 1\sigma$) in derived counts. The background at each
pixel was then estimated using a bi-cubic interpolation between the
coarse grid of backgrounds.
The background subtracted maps were then filtered by the estimated
PSF, including inverse variance weighting, where the noise for each
pixel was estimated from the noise map. 

The maps from all three bands are then combined with weights set by
the local inverse variance, and also the prior expectation of the SED
of the galaxies. We tried a flat-spectrum prior, where equal weight is
given to each band and also 250$\mu$m weighting, where only the
250$\mu$m was included. At the depth of the filtered maps source
confusion is a significant problem, and the higher resolution of the
250$\mu$m maps outweighed the signal-to-noise gain from adding in the
other bands. So for our current data we chose to use the 250$\mu $m
only prior for all our catalogues. We plan to revisit this issue in
future data-releases,  but also compared the number counts produced by our chosen method and the flat spectrum prior and found no significant differences. Further information on our background subtraction and source extraction methods can be found in Maddox et al. (2010) and Rigby et al. (in preparation).

All local peaks are identified in the 250$\mu$m PSF filtered map as
potential sources, and a Gaussian is fitted to each peak. This
provides an estimate of the position at the sub-pixel level, and an
estimate of the peak value, which gives the best flux estimate for a
point source.  The fluxes in other bands were estimated by using a
bi-cubic interpolation to the position given by the 250$\mu$m map. This extraction method was validated against a range of source fluxes in our simulations, and will be discussed in detail in Rigby et al. (in prep). To
produce a catalogue of reliable sources, we selected only sources that
are detected at the 5-$\sigma$ level in any of the bands.  In
calculating the $\sigma$ for each source, we use the relevant noise
map, and add the confusion noise to this in quadrature. The average
1-$\sigma$ instrumental noise values for the PSF filtered maps are 4, 4 and 5.7mJy/beam
respectively in the 250, 350 and 500$\mu$m bands. We estimated the
confusion noise from the difference between the variance of the maps
and the expected variance due to instrumental noise, and find that the
1-$\sigma$ confusion noise is 5, 6 and 7 mJy/beam at 250, 350 and
500$\mu$m.  The resulting total 5-$\sigma$ limits are 33 36 and
45mJy/beam. It should also be noted that overall flux calibration uncertainties
are at the 15\% level for SPIRE (Swinyard et al., in prep).

\section{The H-ATLAS Number Counts}

We use our extracted catalogs as the basis for the determination of the H-ATLAS number
counts in the SPIRE bands. Additionally, those objects identified with optical galaxies whose size extends beyond the
instrumental beam, have fluxes from an explicit aperture extraction with size equal to the instrument beam added in quadrature to the measured object size.
One further source, a merger of two galaxies where the optical radius is too small to include the second component, had an aperture assigned by hand. An additional very bright, extended source, known to be galactic in origin (Thompson et al., in prep), was excluded from our analysis. The resulting catalog of source fluxes in the three bands is cut so that only sources detected at 5$\sigma$ or above in a given band are used for the counts.

Completeness, reliability and the effects of both Õflux boostingÕ (eg. Coppin et al. 2006) and blending were assessed using simulations of our observations. These have the same noise properties as the processed maps
and include a realistic cirrus background, based on IRAS measurements.
Simulated sources were generated using the model of Negrello et al
(2007), with an additional reduction in flux density of 15\% to provide
better agreement with the measured counts. The source extraction method
described above was applied to these images, and the resulting
simulated catalogue was also cut at 5$\sigma$ in each band. 
Flux correction factors matching input and output counts, and accounting for the effects of both boosting/blending and incompleteness, were derived from the simulations. Where greater than 5\%, these corrections were applied to the real counts to produce our final numbers. The band mainly affected by this is the 500$\mu$m channel where flux boosting and blending are most significant due to the large beams. There is also a correction to the lowest flux bin of the 250$\mu$m band as a result of incompleteness.



The Euclidean normalised number counts derived from this data and using the corrections shown in Table 1 are shown in Fig. 1, alongside a variety of models and data from BLAST (Patanchon et al., 2009) and, at 350$\mu$m, Khan et al. (2007). Table 1 also provides raw counts, normalised differential counts and integral counts.

\begin{table}
\tiny
\begin{tabular}{ccccc} \hline
Flux&No. of&Flux&Int. Counts N($>$S) &Diff. Counts\\
(mJy)&Galaxies&Correction&per sq. deg.&gal. sr$^{-1}$Jy$^{1.5}$\\ \hline
\multicolumn{5}{c}{250$\mu$m}\\ \hline
      800&	10&		1  &0.7$\pm$0.2& 2200$\pm$700\\
     450&	4&		1  &1$\pm$0.3&1200 $\pm$600\\
      350&	14&		1  &2$\pm$0.4&2300$\pm$600\\
     250&	16&		1  &3$\pm$0.5&1100$\pm$300\\
     175&	25&		1  &5$\pm$1.0&1500$\pm$300\\
     125&	115&	1 &10$\pm$1.0&2900$\pm$300\\
     93&	 69&		1  &20$\pm$1&4200$\pm$500\\
    83&	126&	1  &25$\pm$2&5700$\pm$500\\
    73&	 232&	1  &45$\pm$2&7600 $\pm$500\\
    63&	418&	1  &70$\pm$2&9600 $\pm$500\\
    56&	345&	1  &100$\pm$3&12000 $\pm$650\\
    52&	506&	1  &150$\pm$5&14000$\pm$600\\
    47&	710&	1  &200$\pm$5&16000$\pm$600\\
    42&	1093&	1  &300$\pm$5&18500 $\pm$600\\
    38&	1702&	1&374$\pm$5&21500 $\pm$500\\
    34&	898&	1.1&436$\pm$6&23000$\pm$800\\ \hline		
\multicolumn{5}{c}{350$\mu$m}\\ \hline
 700&	7&	1&0.5$\pm$0.2&800$\pm$300\\
 250&	8&	1&1.0$\pm$0.3&600 $\pm$200\\
 175&	8&	1&1.6$\pm$0.3&500$\pm$200\\
 125&	27&	1&3.5$\pm$0.5&700$\pm$150 \\
 91&		16&	1&4.6$\pm$0.5&900$\pm$250\\
 82&		43&	1&7.6$\pm$1&2000$\pm$300\\
 71&		92&	1&14.0$\pm$1&2800$\pm$ 300\\
  61&195&	 1&28$\pm$1&4200$\pm$300\\
 55&	158&	1&39$\pm$2&5000$\pm$ 400 \\
 50&	256&	1&56$\pm$2& 6600$\pm$400\\
46&	458&	1&88$\pm$3&9400$\pm$ 450\\ 
41&	669&	1&134$\pm$3&11000$\pm$400\\
38&	594&	1&176$\pm$4&13000$\pm$550\\ \hline
\multicolumn{5}{c}{500$\mu$m}\\ \hline
250	&4	&	1&0.28$\pm$0.15&285$\pm$150\\
150	&7	&	1&0.76$\pm$0.25&140$\pm$50\\
 95	&3	&	1&0.97$\pm$0.3&190$\pm$100\\
 82&13	&	1&1.88$\pm$0.4&370$\pm$100\\
 70&26	&	0.89&3.68$\pm$0.5&580$\pm$100\\
 63&16	&	0.84&4.79$\pm$0.6&580$\pm$150\\
 60&19	&	0.82&6.11$\pm$0.7&760$\pm$200\\
 56&43	&	0.82&9.1$\pm$0.8&1100$\pm$200\\
 52&76	&	0.83&14.4$\pm$1.0&1300$\pm$150\\
 48&50	&	0.84&17.8$\pm$1.0&1900$\pm$300\\ \hline
\end{tabular}
\small
\caption{H-ATLAS Galaxy Counts. Correction column gives the effective flux correction factor required to match the output to 
input counts from the simulations, and the fluxes quoted are after correction.}
\end{table}

\begin{figure}
\includegraphics[angle=0,width=9cm,height=6.5cm]{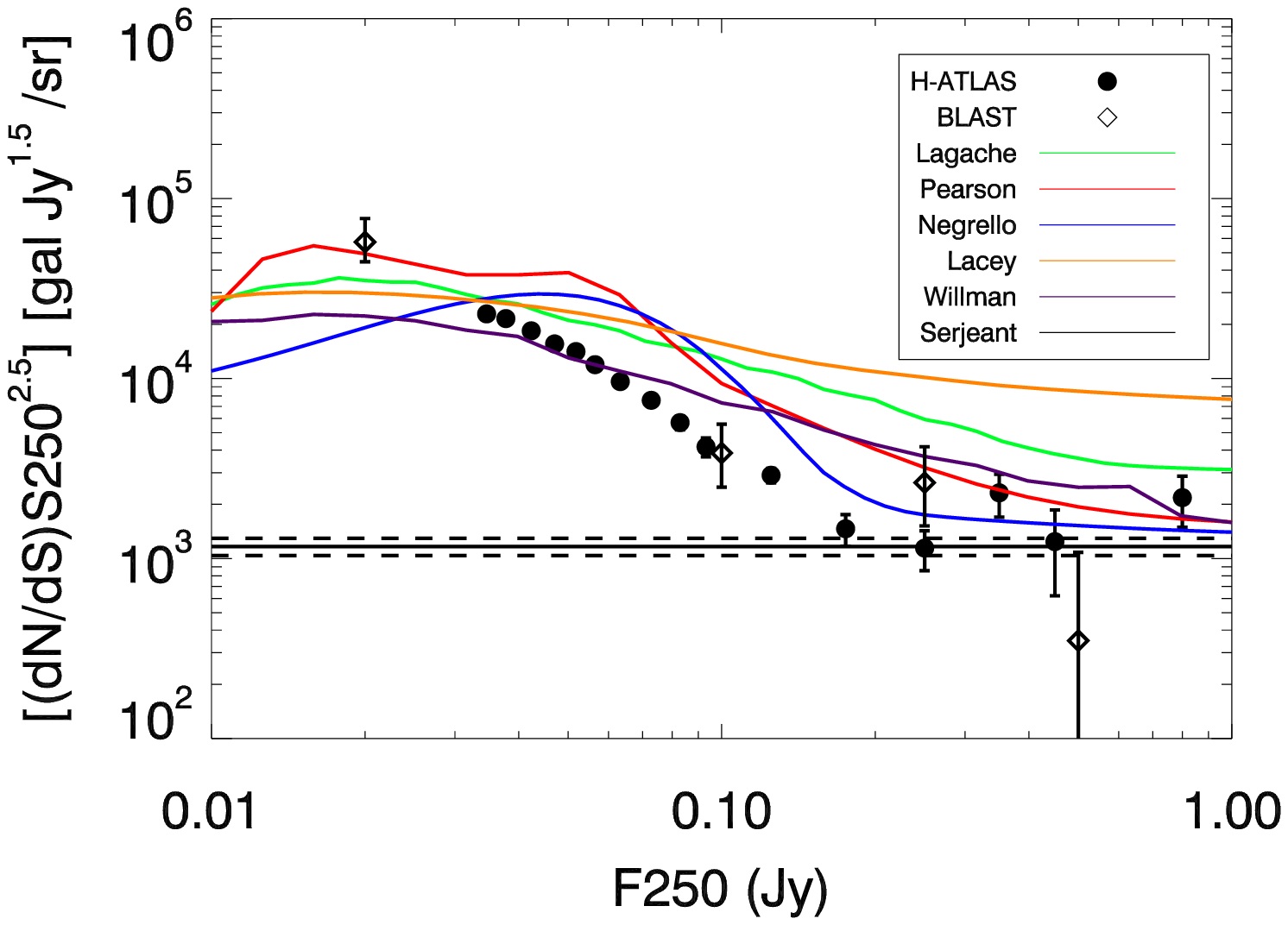}
\includegraphics[angle=0,width=9cm,height=6.5cm]{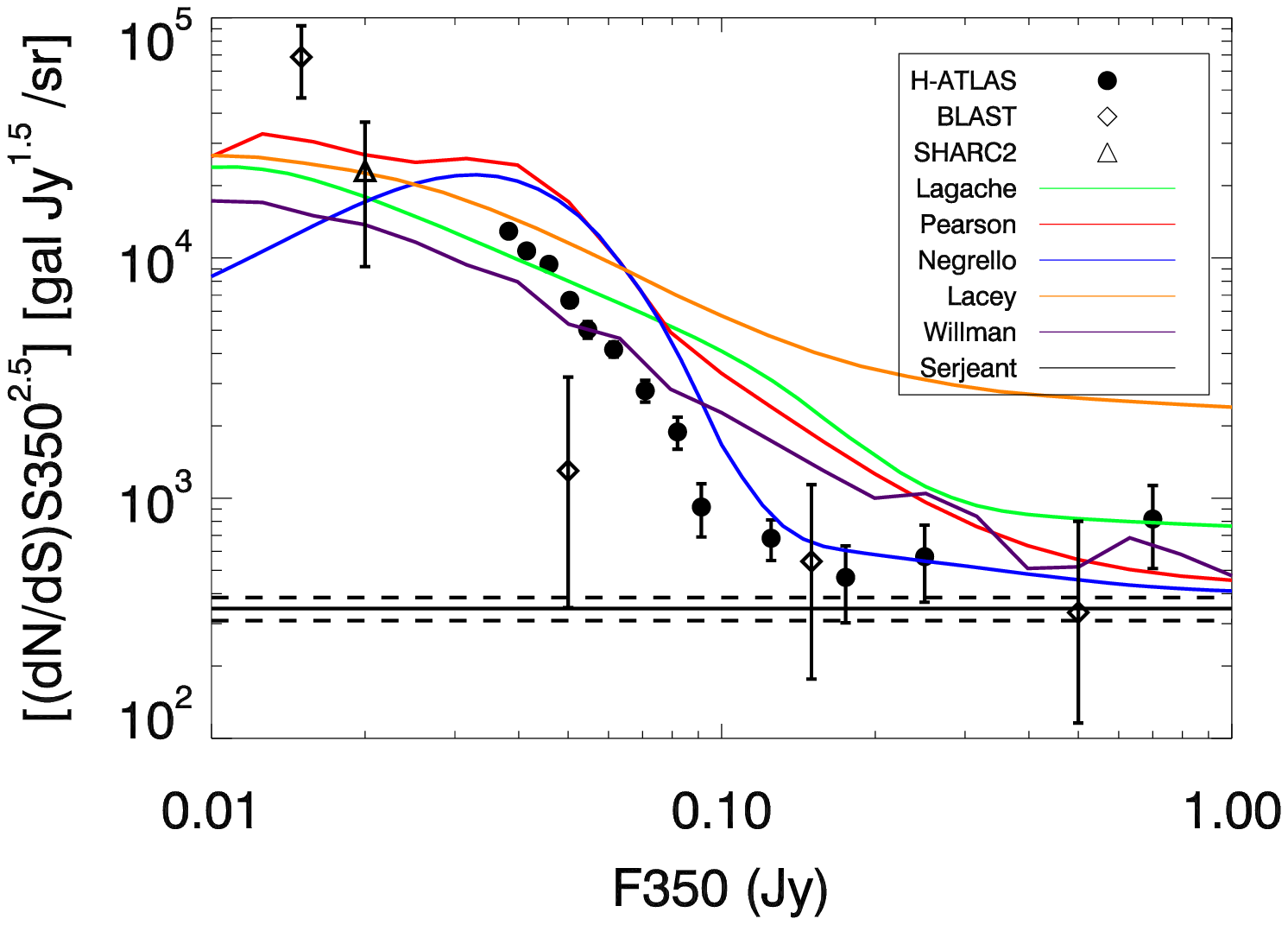}
\includegraphics[angle=0,width=9cm,height=6.5cm]{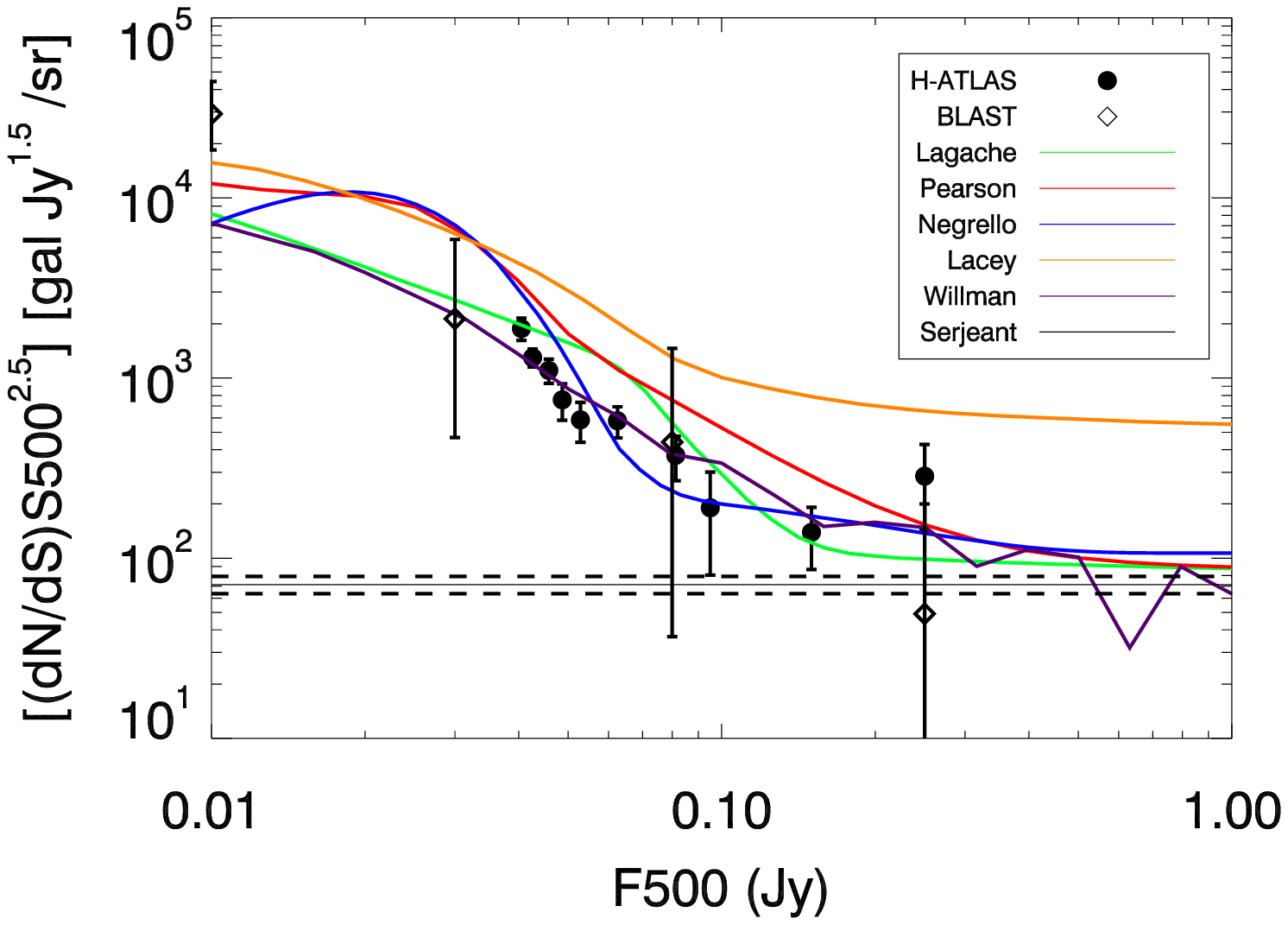}
\caption{H-ATLAS Euclidean Normalised Differential Number Counts compared to models and other data. Data from this work are shown as solid dots. Data from BLAST is shown as open diamonds. The solid horizontal line shows the asymptotic local differential counts from Serjeant \& Harrison (2005) with dashed lines showing the $\pm$11\% fractional uncertainly. Models are shown as follows: Lagache et al. (2004) - green; Pearson \& Khan (2009) - red; Negrello et al. (2007) - blue; Lacey et al.(2008) - orange; Wilman et al. (2010) - purple. {\bf Top:} 250$\mu$m counts. {\bf Middle:} 350$\mu$m counts, with the 350$\mu$m point of Khan et al. (2007) shown as an open triangle. {\bf Bottom:} 500$\mu$m counts.}
\end{figure}

\section{Discussion}

We compare the H-ATLAS counts to a range of models and to existing data from BLAST (Patanchon et al., 2009), SHARC2 (Khan et al., 2007) at 350$\mu$m, and the Serjeant \& Harrison (2005) study of the local far-IR/submm luminosity function. The models include those of Lagache et al (2004), Pearson et al. (2009), Negrello et al. (2007), Lacey et al., (2009) and Wilman et al. (2010) and cover a wide range of modelling techniques and assumptions. The Lagache et al. (2004) model is a backward-evolution model in which two template SEDs, one for normal galaxies and another for starbursts, are evolved independently from z=0. The Pearson et al. (2009) model is a similar backward-evolution model but instead uses seven luminosity dependent template SEDs. The rate of evolution is also luminosity dependent with stronger evolution in the more luminous sources (bright mode evolution in the terminology of Pearson et al. (2009)). The Negrello et al. (2007) count model includes contributions from a range of populations. These include submm galaxies evolving according to a physical model (Granato et al., 2004), radio sources, as modelled by DeZotti et al. (2005), and late-type (starburst plus normal spiral) galaxies. Unlike the others, the counts of SMGs in this model also include a contribution from strongly-lensed galaxies,  quantified following the recipes of Perrotta et al. (2002). The Lacey model is calculated from a theoretical model of galaxy formation based on CDM (Lacey et al. 2010). It uses the GALFORM semi-analytical galaxy formation model in combination with the GRASIL model for the reprocessing of starlight by dust (Granato et al 2000). The galaxy formation model is the same as that described in Baugh et al (2005), and compared to Spitzer data in Lacey et al (2008), and incorporates a top-heavy IMF in starbursts triggered by galaxy mergers. The Wilman et al. (2010) source counts are based on the large SKADS radio continuum simulation (Wilman et al. 2008) and uses the radio-FIR relation together with a model of the evolution of this relation, along with various prescriptions to account for the AGN and starburst populations, to predict the counts in the FIR/submm. All of these models are broadly consistent with existing data on galaxy counts, redshift distributions and luminosity functions from the mid-IR to radio bands, including data from IRAS, Spitzer and ground-based submm surveys.

A clear feature in the observed counts is a steep increase above an extrapolation of the bright counts to fainter fluxes. This increase starts at fluxes $\sim$100---200mJy. Despite being hinted at in earlier observations, Herschel allows this rise in counts to be unambiguously detected and studied in detail for the first time. The feature is seen in all three SPIRE bands, but is most pronounced in the longest wavelength, 350 and 500$\mu$m, bands  where the counts rise roughly twice as much as at 250$\mu$m. Such a feature is predicted in a number of models though no model is a particularly good fit to the counts in all bands. The best performing model with respect to the shape of the counts is that of Negrello et al. (2007), but this model overpredicts the effect of the bump at 250 and 350 $\mu$m. The Negrello model also provides a better fit to the brighter counts than most of the other models, especially at 350$\mu$m, but there is clearly much work to be done to improve the match between any of these models and the observed data. The number count rise represents the strongly evolving luminous ($> 10^{11} L_{\odot}$) galaxy population that is a generic feature in all galaxy evolution models that are able to fit the submm galaxies. The details of this evolution, possibly combined with other effects such as gravitational lensing (a feature of the Negrello model), control the shape and position of the rise. Its strength as a function of wavelength will be dependent on the detailed redshift and possibly evolving SEDs of the rapidly-evolving population which can also be constrained by the colour of individual sources (see eg. Amblard et al. 2010). Its prominence at 350 and 500$\mu$m compared to 250$\mu$m thus argues that the sources responsible lie at z$>$1.5, as suggested by Granato et al. (2004). If they lie at lower redshifts they would require colder SEDs. It is also interesting to note that some models (eg. that of Lacey) appear to be strongly excluded by the observations and that the largest discrepancy between predicted and observed counts is at the bright end, where the full H-ATLAS survey will be uniquely powerful. These models are also unable to reproduce the asymptotic local counts from Serjeant and Harrison (2005) based on IRAS data. This suggests that our knowledge of brighter, more local far-IR/submm galaxies is not as good as we might wish.

These new counts from H-ATLAS set the scene for a much better understanding of the enigmatic far-IR/submm population. The SDP data presented here represents less than 3\% of the total area that will be covered by H-ATLAS. 
Although this is only a small fraction of H-ATLAS, these data already include more 
than 5 times as many ($>5\sigma$) sources as previously published at these 
wavelengths. As further data is added, more detailed measurements towards the bright end will allow the effects 
of strong lensing and local sources on the counts to be determined.
H-ATLAS is also large enough to totally overcome the effects of cosmic variance, 
and allow it to be studied by comparing counts in sub-regions.

\section{Conclusions}
We have used the first $\sim$14 sq. deg. of the H-ATLAS project observed by Herschel to determine the number counts of far-IR/submm sources in the SPIRE bands at 250, 350 and 500$\mu$m. We find that there is a steep rise in the counts at fluxes of 100-200mJy. These observations are compared to a range of galaxy evolution models and data from BLAST and elsewhere. The H-ATLAS counts are found to be consistent with previous observations, but no theoretical model is an ideal fit. Some of the models are quite poor matches to the observations, with counts at fluxes brighter than $\sim$200mJy setting strongly excluding some models. The best fitting model of those considered is that of Negrello et al. (2007). These observations represent just 3\% of the total area to be covered by the H-ATLAS survey. Once completed this project will clearly be a very powerful tool for understanding the far-IR/submm galaxy population. 

     

\end{document}